\magnification=1200
\tolerance=10000
\hsize 14.5truecm
\hoffset 1.25truecm
\font\cub=cmbx12
\font\ninerm=cmr9
\baselineskip=24truept
\parindent=1.truecm
\def\ref{\par\noindent\hangindent 20pt}

\def\mincir{\raise -2.truept\hbox{\rlap{\hbox{$\sim$}}\raise5.truept
\hbox{$<$}\ }}
\def\magcir{\raise -2.truept\hbox{\rlap{\hbox{$\sim$}}\raise5.truept
\hbox{$>$}\ }}
\def\asymp{\raise -4.3truept\hbox{$ \ \widetilde{\phantom{xy}} \ $}}

\input newsym

\vskip 3 cm
\centerline{{\cub Some remarks on the question of}}

\centerline{{\cub charge densities in stationary--current--carrying 
conductors}}
\vskip 5 mm
\centerline{{\bf L. Baroni}, \footnote {$^1$} {\baselineskip=14
truept {\ninerm Department of
Physics, University of Bologna and INFN Sezione di Bologna, Italy}}
 ~{\bf E. Montanari},
\footnote {$^2$} {\baselineskip=14
truept {\ninerm Department of
Physics, University of Ferrara and INFN Sezione di Ferrara, Italy; 
E--mail address: montanari@fe.infn.it}}
 ~{and} 
 ~{\bf A.D. Pesci} {$^1$}}
\vskip 1.5truecm
\noindent
{\bf Abstract.} Recently, some discussions arose as to
the definition of charge and the value of the density of charge
in stationary--current--carrying conductors. 
We stress that the problem of charge definition comes from
a misunderstanding of the usual definition. 
We provide some theoretical elements which suggest
that positive and negative charge densities are equal
in the frame of the positive ions.
\vskip 1 truecm

\vskip 2truecm
\centerline{Nuovo Cimento B {\bf 109}, 1275 (1994)}
\vfill\eject

\noindent
{\bf 1. Introduction}
\medskip

Recently Ivezi\'c [1--4] has questioned the correctness of the usual 
charge definition and raised the problem of the charge density inside 
a current--carrying conductor. He gives a new definition of 
charge and suggests that the ion charge 
density, as measured in the ions reference frame, could be equal to 
the electron charge density as measured in the electrons reference frame. 
This is contrary to the usual view that these two charge densities are 
equal in the ions reference frame.
A consequence of Ivezi\'c's assumption is that outside a conducting 
wire there should be a static electric field when a stationary 
current is flowing in the conductor.
This author also claims that in the seventies some 
experiments were carried out [5--7] (actually also recently in [8,9]) 
which are in agreement with this 
new definition of charge.
As a consequence of Ivezi\'c's papers, many authors [10--13] have 
taken up a definite position against his ideas which provoked a 
discussion which, in our opinion, has not yet settled down the question.
The paper is organized as follows. In sect. 2 we show that the usual 
charge definition is the correct 
one; in sect. 3 we show how to deal correctly with 
invariant integrals, while in sect. 4 we disprove Ivezi\'c's 
hypothesis of the non--zero 
electric field outside current--carrying conductors.

\bigskip
\noindent
{\bf 2. Independence of the charge on the velocity}
\medskip

Purcell [14] had pointed out that the independence of the charge from 
the velocity means that the integral
$$
{1\over 4\pi}\ \int_{A(t)}{\bmit E\ d\bmit a} = Q
\eqno (2.1)
$$
does not depend on the motion of the particles inside the closed 
surface $A(t)$; moreover, if we choose another closed surface which 
contains the same number of particles, then 
the value of the flux of the electric field from this surface is the 
same. If this happens for an inertial reference frame then, because of 
the principle of relativity, it must be true for any other inertial frame. 
Therefore, if we consider in a frame $O$ a system of charges at the time 
$t$ in a volume surrounded by the closed surface $A(t)$ and in a frame 
$O'$ in motion with constant velocity with respect to $O$ any volume 
which contains at the time $t'$ the same particles surrounded by the 
closed surface $A'(t')$, then
$$
\int_{A(t)}{\bmit E\ d\bmit a} = \int_{A'(t')}{\bmit E'\ d\bmit {a'}} 
\eqno (2.2)
$$ 
where the two integrals are evaluated at the times $t$ and $t'$ 
respectively.
We want to point out that  eq. (2.2) does not give a recipe to 
obtain the closed surface $A'(t')$ nor the time at which to perform 
the integration once $A(t)$ has been given. Only for a 
closed system inside a surface $A$
(here with closed system we mean that no charges cross its 
boundaries) it is correct to take the surface, for instance, as 
simultaneously seen in $O'$ and perform the integration in any 
instant of time $t'$. In this case in fact 
the problem is independent of time.
On the contrary, if we consider a non--closed system, 
like a piece of wire with a steady current, then it is not trivial to 
choose a correct surface in $O'$ (a possible wayout of this 
difficulty has been proposed in [10]).

Purcell in his definition of invariance of charge does only 
state that equation (2.2) must hold if, and only if, the boundaries in 
the two reference frames contain the same particles. And this 
is equivalent to admit the postulate that the charge of a 
particle does not depend on its motion.
These specifications are to show the right way in which eq.
(2.2) should be interpreted. 
In fact we think that in [1, 4] there has been a misunderstanding 
about the meaning and validity of equation (2.2). 

We want to stress that the invariance of the charge as defined in 
(2.2),
{\it i.e.} its independence from motion, 
is a consequence of the fact that the equation of charge 
conservation is a continuity equation 
$$
{\partial \rho\over \partial t} + div{\bmit j} = 0
\ ,
\eqno (2.3)
$$
fulfilled in every inertial frame.

To show this let us consider a charge at rest in the origin of an 
inertial frame 
$O$. At the time $t=0$ we turn on a suitable field of force which 
causes an acceleration along the $x$ direction for a time $\tau$. 
Therefore, at the time $t=\tau$ the charge is moving with a uniform 
velocity $V$ in the $x$ direction. Now let us consider a closed 
spherical surface with centre at the origin and radius $r_0$ large 
enough not to be crossed by the charge for an interval of time 
$I=[\tau,T]$ (where $T>\tau$). If we integrate eq. (2.3) in 
this reference frame  
on the volume enclosed by that surface for every $t < T$ we obtain
$$
{dQ\over dt} = {d\over dt} \int{\rho dV} 
= - \int{\bmit j\ d\bmit a} = 0
\eqno (2.4)
$$
This means that the total charge inside a close surface is not 
affected by the state of motion but can vary if and only if some 
charged particles cross the boundary.

This reasoning can be done for 
any number of charges and any initial configuration. 
Therefore (2.2) (in the sense and with the specification we 
have made) derives from the assumption of equation (2.3) for the 
conservation of charge.

All these considerations show that a
correct way to think about charge conservation is 
the continuity equation.
According to this equation, it is not important that the charge in a 
certain volume does not change, but what matters is the fact 
that the change should be only due to the charges 
that cross the closed surface which is the boundary of that volume. 

\bigskip
\noindent
{\bf 3. The invariant integral of charge}
\medskip

If we want to consider a relativistic scalar quantity, that 
is something which is invariant under Lorentz transformation, 
eq. (2.2) becomes useless first because the integrand is not 
written as a relativistic scalar and second because there is no a 
priori relation between $A(t)$ and $A'(t')$. 
One thing is to say that equation (2.2) holds and another one is to  
construct an invariant quantity.
The first statement means that the charge does not depend on the 
velocity.
The second means that if we want to find, in a relativistic 
invariant way, the 
same amount of charges in two different reference frames we must also 
take into account the fact that the difference of simultaneity gives 
rise to differences as to the charges contained contemporaneously in 
a volume [15].
To obtain such an invariant quantity we must consider as 
integrand a relativistic scalar and a hypersurface as a domain. In 
this way if one changes the 
inertial frame, the value of the integral remains the same, but its meaning 
in the two frames is in general different because of the difference 
of simultaneity.
As pointed out by [11] the correct invariant quantity is that given in 
[16], that is  
$$
Q = {1\over c} \int_H{j^\mu dS_\mu} 
\eqno (3.1)
$$
where $H$ is a hypersurface.
In this way $Q$ turns out to be the sum of those charges
the world lines of which cross this hypersurface.
When one has to do with integrals over hypersurfaces, the way to handle 
them is to parametrize the domain. In the four--dimensional space--time 
of special relativity this means that we must consider the coordinates 
as functions of three real parameters, that is to say
$$
\eqalign {
x^\mu&=\phi^\mu(u_i)\qquad\qquad u_i\in U_i\subset \Re
\cr
i&\in {1,2,3} 
\cr}
\eqno(3.2)
$$
Then by definition
$$
\int_H{j^\mu(x^\nu) dS_\mu} = 
\int_{U_1}{du_1}\int_{U_2}{du_2}\int_{U_3}{du_3}\ 
j^\mu(u_i)\ n_\mu 
\eqno(3.3)
$$
where (putting $\phi=(\phi^0,\phi^1,\phi^2,\phi^3,)$ and 
$\phi_{u_i}=\partial \phi / \partial u_i$),
$$
n^\mu = (\phi_{u_1}\wedge\phi_{u_2}\wedge\phi_{u_3})^\mu = 
- {1\over 6} e^{\mu\alpha\beta\gamma} D_{\alpha\beta\gamma} 
\eqno (3.4)
$$
and
$$
D^{\alpha\beta\gamma} =  det
\left(
\matrix {\phi^\alpha_{u_1} & \phi^\beta_{u_1} & \phi^\gamma_{u_1} \cr
\phi^\alpha_{u_2} & \phi^\beta_{u_2} & \phi^\gamma_{u_2} \cr
\phi^\alpha_{u_3} & \phi^\beta_{u_3} & \phi^\gamma_{u_3}}
\right)
\eqno(3.5)
$$
If we perform a Lorentz transformation, then one has 
$$
\eqalign {
x'^\mu=l^\mu_\nu x^\nu =&\ l^\mu_\nu \phi^\nu (u_i) = \psi^\mu (u_i)
\cr 
j^\mu[\phi^\nu(u_i)] n_\mu (u_i) &=
j'^\mu[\psi^\nu(u_i)] n'_\mu (u_i)
\cr}
\eqno(3.6)
$$
In this way we have that (putting 
$x'=(x'^\mu)=(l^\mu_\nu x^\nu)=\bmit l x$)
$$
\eqalign {
\int_H{j^\mu(x^\nu)}&\ dS_\mu = 
\int_{U_1}{du_1}\int_{U_2}{du_2}\int_{U_3}{du_3}\ 
j^\mu[\phi(u_i)]\ n_\mu =
\cr 
&= \int_{U_1}{du_1}\int_{U_2}{du_2}\int_{U_3}{du_3}\ 
j'^\mu[\psi(u_i)]\ n'_\mu =
\int_{\bmit l(H)}{j'^\mu(x'^\nu)\ dS'_\mu} 
\cr }
\eqno(3.7)
$$
so that integral (3.3) is invariant under Lorentz transformation.

Another way to consider integral (3.1) is by means of differential 
forms. In fact let us consider $j_\mu$ as the components of the 
1--form
$$
\bmit J = j_\mu \ dx^\mu
\eqno (3.8)
$$
From (3.8) we can define the dual 3--form $\bmit {^*\! J}$: 
$$
\eqalign {
\bmit {^*\! J} = {1\over 6}&\ j_{\alpha\beta\gamma} 
 dx^\alpha\wedge dx^\beta\wedge dx^\gamma
\cr
&j_{\alpha\beta\gamma} = e_{\alpha\beta\gamma\mu} j^\mu
\cr }
\eqno (3.9)
$$
where $e_{\alpha\beta\gamma\mu}$ is the totally antisymmetric tensor 
as defined, for instance, in paragraph 6 of ref. [16].
With these definitions integral (3.1) can be written as (see, 
for instance, Box 4.1 D in ref. [17])
$$
Q = \int_{H} {\bmit {^*\! J}}
\eqno (3.10)
$$
If we consider the exterior derivative of $\bmit {^*\! J}$ we see 
that it is zero because of the continuity equation (2.3); that is to 
say $\bmit {^*\! J}$ is the exterior derivative of a 2--form. This 
2--form is the dual of the tensor of the electromagnetic field 
$\bmit F = F_{\mu\nu} dx^\mu\wedge dx^\nu$:
$$
\bmit {d ^*\! F} = {4\pi\over c} \bmit {^*\! J}
\eqno (3.11)
$$
This is very important in the calculation of (3.10). In fact this 
implies (as in the one--dimensional case when one deals with an exact 
1--form) that the value of the integral is left unchanged when, in 
order to simplify the calculatins, we deform the hypersurface of the 
domain leaving the boundaries unaltered.

This way to perform calculation is equivalent to the one 
used in [11].
The only difference is that Bili\'c in [11] considers a two dimensional 
space--time. In this particular situation the integrand is an exact 
differential form (because of the continuity equation) and so the 
value of the integral is the same for any path of integration between 
the points $A$ and $B$ which must be the same contrary to 
what has been said in [4].
In other words, this means that in [4] Ivezi\'c does not interpret 
correctly eq (3.1).

The way in which we have considered integral (3.1) implies automatically 
that we are dealing with the same total carge as well with 
the same charged particles in the two reference frames.
The root of this lies in the continuity equation. In fact, on the
one hand, one has to remember (as pointed out by Ivezi\'c in [4]) that 
$j^\mu$ is a four--vector because of the postulate of relativity that 
imposes the covariance of continuity equation. In this way 
$j^\mu\ dS_\mu$ is a scalar (and therefore we are dealing 
with the same charged particles). On the other hand, if we 
have the same particles, the continuity equation implies 
that we have also the same total charge, independent on 
their state of motion.

In this way we have stressed once more that, as pointed out by 
paragraph 29 of ref. [16] and Box 4.1 D of ref. [17], the 
invariance of charge, stated 
by equation (3.1), is strictly related to the continuity equation 
of charge (2.3).

\bigskip
\noindent
{\bf 4. The non--zero electric field hypotesis}
\medskip

Even if in [1] 
the questions of the exact way to interpret the charge invariance
and of the existence of an $\bmit E \neq 0$  externally
to a current--carrying conductor may seem to be related,
actually they are not (as stressed in [4]).

Historically, the existence of an $\bmit E \neq 0$  outside
a current--carrying conductor was largely discussed in the 
literature (see, for instance, [5] sect. {\bf 1}).
Even if the usual belief is that 
such an $\bmit E$ does not exist,
however there is no experimental evidence.
This question is not settled in the framework
of the usual Maxwell theory (where the charge is assumed
not to depend on the velocity); it can in fact happen
that the positive charge density in a flowing
current turns out to be different from that of
negative charges.

According to [1] the problem of the existence of this field 
can be traced back to that of knowing in which frame 
$\lambda_+ = \lambda_-$.
In [1,4] and [18] it is clearly stressed
that two physical hypoteses are relevant to this point.
The first one assumes that 
$\lambda_+ = \lambda_-$ in the wire frame,
and this gives
$\bmit E = 0$
according to the common belief;
the second one assumes that 
$\lambda_+$ (as evaluated in the ions rest frame)
is equal to
$\lambda_-$ (as evaluated in the electrons rest frame).
Ivezic prefers the latter
because the charges are treated in a symmetric way.
He also believes that some experimental results confirm it [5].

According to Ivezi\'c view one expects a radial field
$E = \delta\lambda/(2\pi r)$ where $\delta\lambda$ is the 
absolute value of the difference of charge density in the 
rest frame of the lattice and $r$ is the radial distance 
from the wire.
It turns out that
$$
\delta\lambda = (\gamma -1)\lambda_+ \simeq {1\over{2}}
\left ({v\over c}\right )^2 \lambda_+
\eqno(4.1)
$$
where $v$ is the drift velocity and, as usual,
$\gamma = 1/\sqrt{1-{v^2}/c^2}$.
The field $\bmit E$ is a second order quantity in $v/c$
and therefore extremely small under ordinary conditions.
This fact prevented from verifying its existence until now.  
One can hope that experiments with superconducting materials
like those reported in [5] can settle the question.
Such experiments have been however proposed not to this end 
but to measure possible 
second--order deviation from Maxwell's equations due to 
an hypothetical dependence of the charge on the velocity; in 
fact it was taken for granted by these experimentalists that 
ions and electrons charge 
densities where the the same in the rest frame of the ions.
This same assumption was made in the paper by Baker [19] who 
has shown, starting from Lienard-Wiechert potentials, that there 
is no electric field produced by the charge drift in a 
current--carrying conductor in accordance with Gauss's law.

Ivezi\'c in [1] and [4] raised the question of 
the possible existence of $\bmit E \neq 0$ outside a  
current--carrying conductor in the 
framework of Maxwell's equations without postulating a 
charge dependence on the velocity as proposed in [5] but 
assuming $\lambda_+ \not= \lambda_-$ in ions frame. 

In our opinion there are at least three reasons why the two 
densities should be equal in the ions frame.

The first one is related with thermal motion. 
Already in [12], Singal asked why Ivezi\'c did not take 
into consideration the effects of thermal noise of electrons.
In his words this means that ``it is not clear how in Ivezi\'c's approach 
the effects of the thermal velocities of electrons, 
many orders of magnitude larger than their drift 
velocities, could in some unambiguous way be incorporated or 
at least shown to cancel, since his derived electric fields
(see eq. (4.1)) depend upon the square of the velocity of 
moving charges."
It is clear that if we consider an isolated conductor its charge 
cannot vary. But let us consider the case of a wire that connects two 
charge reservoirs. Let us suppose that the distribution of the 
velocity of electrons at a certain temperature $T$ (ions are 
considered at rest because of their large mass) is given by a 
function $f_T(v)=f_T(-v)$ normalized to unity 
(that is to say $\int_\Re {}f_T(v) dv =1$). Let us call $O(v)$ 
the reference frames in motion with  velocity $v$ with respect to ions. 
If Ivezi\'c ideas are correct, in that frame the charge density of 
electrons with velocity lying in the interval $(v, v+dv)$ must be 
equal to that of the ions in $O(0)$ with the opposite sign. 
This means that in the ions rest frame these electrons are 
characterized by a linear charge density 
$\lambda (v) = - \gamma (v) \lambda_+$.
In this way the linear negative--charge density would be given by
$$
\Lambda (T) = \int_{-\infty}^{+\infty} {}\lambda (v) f_T(v) dv   
\eqno (4.2)
$$
where $\Lambda$ depends only on the temperature $T$. 
This means that there is an excess of negative charge per unit length 
of the amount $\delta \lambda (T) = \lambda_+ + \Lambda (T)$.
If we consider ordinary temperature, the gas of electrons in the 
conductor is almost completely degenerate. In this way we can give an 
extimation of $f_T(v)$ independent of the temperature. To get the 
order of magnitude we can put
$$
f(v) = {1\over 2 v_F} \qquad v\leq |v_F|
\eqno (4.3)
$$
where $v_F$ is the Fermi velocity.
In this way up to the second order in $(v_F/c)$
$$
\Lambda \simeq - 
\left [1 + {1\over 6} \left ({v_F\over c}\right )^2\right ] 
\lambda_+.
\eqno (4.4)
$$
The linear charge excess is then given by 
$$
\delta \lambda \simeq 
- {\lambda_+\over 6} \left ({v_F\over c} \right )^2.
\eqno (4.5)
$$
For a centimeter of copper wire with a cross section $S=10^{-4}\ cm^2$, 
taking into account that $v_F = 1.56 \times 10^8\ cm/sec$ and
$\lambda_+ = 8.5 \times 10^{22} e S\ C/cm$ (where $e$ is the absolute 
value of the electron charge) one 
has a charge $Q = 6 \times 10^{-6} C$! 

A second way to look at the problem of charge density is to consider 
the current flowing in the wire as due to an acceleration of electrons 
at rest in the wire caused by an applied electric field. The 
steady state is reached because of the inner resistance of 
the wire.
Now all the electrons undergo the same accelerating field 
and therefore, assuming the same initial velocity (in this 
case equal to zero), their distance cannot change in the 
ions frame (cf. chapt. 20 in ref [16]). This means by the way that 
during the acceleration the proper distance among electrons 
increases due to the Lorentz contraction.

As a third consideration about Ivezi\'c problem, we take into account 
Ohm's law.
In the interior of a conductor at rest one has
$$
\bmit j = \sigma \bmit E
\eqno (4.6)
$$
where $\sigma$ is the conductivity.
By means of Gauss's law $div \bmit E = 4\pi \rho$ and the continuity 
equation, one easily finds an equation for the density of the 
charge inside a conductor
$$
{\partial \rho\over \partial t} + 4\pi\sigma\rho = 0
\eqno (4.7)
$$
The solution of this equation is
$$
\rho = \rho_0 exp \left (-{t\over \tau} \right )
\eqno (4.8)
$$
where $\tau = (4\pi\sigma)^{-1} \simeq 10^{-18}\ sec$ that is a time 
which is correct to take equal to zero in a macroscopic theory.
This implies that conduction electrons inside a conductor have always
the same density as ions.
Moreover, for steady currents 
$0=div \bmit j = 4\pi \sigma\rho$, that is to say $\rho=0$.
Therefore, Ohm's law requires that inside a conducting wire 
the two charge densities are equal in the reference frame of 
ions. A possible charge excess, as that assumed 
by Ivezi\'c, can only stay on the surface and this destroys 
the symmetry reason invoked by him.

Therefore, it appears reasonable that the birth of a current 
does not modify the electron charge density in the frame in 
which ions are at rest (and in which also the electrons 
were at rest before an electric field was applied).

In conclusion, even though Ivezi\'c has the merit to have 
stressed that both situations (i.e. charge densities equal 
in the frame of the wire or in their own reference frame) 
are compatible with Maxwell's equation, the above 
considerations (which lie outside the domain of this theory) 
show however that the first of the two 
situation envisaged, {\it i.e.} the one commonly assumed, is the 
right one.

\bigskip
\noindent
{\bf 5. Conclusions}
\medskip  
 
We have shown that the question raised by Ivezi\'c about the 
non--invariance of the charge under Lorentz trasformations was 
originated by a misunderstanding of the usual charge 
definition. If such definition is properly understood, one 
has the usual charge invariance and there is no necessity of 
looking for new invariant charges as claimed by Ivezi\'c.

As to the ambiguity pointed out by Ivezi\'c about the reference frame 
in which the positive charge density should be equal to the negative 
one in a conductor carrying a stationary current, we have shown that 
the usual belief, that is to say that $\lambda_+=\lambda_-$ in ions 
rest frame, is the correct one. In fact, if one assumes Ivezi\'c's 
idea, the following three facts, clearly in contrast with experimental 
results, should happen: 
1) there should be present very big effects due to the 
thermal motion of the electrons; 2) the same accelerating field on the 
same particles with the same initial velocities would change the mean 
distance among them; 3) Ohm's law would be no longer valid.

\vfill\eject
\vskip 1.2truecm
\noindent
{\bf References} 
\bigskip

\ref {[1]~~~T. Ivezi\'c, Phys. Lett {\bf A 144}, (1990) 427.}

\ref {[2]~~~T. Ivezi\'c, Phys. Lett. {\bf A 156}, (1991) 27.}

\ref {[3]~~~T. Ivezi\'c, Phys. Rev. {\bf A 44}, (1991) 2682.}

\ref {[4]~~~T. Ivezi\'c, Phys. Lett. {\bf A 162}, (1992) 96.}

\ref {[5]~~~W. F. Edwards, C. S. Kenyon, D. K. Lemon, Phys. Rev. 
{\bf D 14}, (1976) 922.}

\ref {[6]~~~D. F. Bartlett, B. F. L. Ward, Phys. Rev. {\bf D 16},
(1977) 3453.}

\ref {[7]~~~D. F. Bartlett, J. Shepard, C. Zafiratos, B. F. L. Ward,
Phys. Rev. {\bf D 20}, (1979) 578.} 

\ref {[8]~~~C. S. Kenyon, W. F. Edwards, Phys. Lett. {\bf A 156},
(1991) 391.}

\ref {[9]~~~D. K. Lemon, W. F. Edwards, C. S. Kenyon, Phys. Lett.
{\bf A 162}, (1992) 105.}

\ref {[10]~~~D. F. Bartlett, W. F. Edwards, Phys. Lett. {\bf A 151},
(1990) 259.}

\ref {[11]~~~N. Bili\'c, Phys. Lett. {\bf A 162}, (1992) 87.}

\ref {[12]~~~A. K. Singal, Phys. Lett. {\bf A 162}, (1992) 91.}

\ref {[13]~~~D. F. Bartlett, W. F. Edwards, Phys. Lett. {\bf A 162},
(1992) 103.}

\ref {[14]~~~E. M. Purcell, {\it Electricity and Magnetism}, 
2nd Ed. (McGraw--Hill, New York, N. Y., 1985).}

\ref {[15]~~~R. Becker, {\it Teoria dell'elettricit\`a}, 
Vol. II (Sansoni Edizioni Scientifiche, Firenze, 1950).}

\ref {[16]~~~L. D. Landau, E. M. Lifsits, {\it Course of Theoretical 
Physics, Vol. 2: The Classical Theory of Fields} 
(Pergamon Press, Aberdeen, 1975).}

\ref {[17]~~~C. W. Misner, K. S. Thorne, J. A. Wheeler, 
{\it Gravitation} (Freeman, San Francisco, Cal., 1973).} 

\ref {[18]~~~T. Ivezi\'c, Phys. Lett. {\bf A 166}, (1992) 1.}

\ref {[19]~~~D. A. Baker, Am. J. Phys. {\bf 32}, (1964) 153.} 


\vfill\eject
\bye